\documentclass[aps,prl,twocolumn,groupedaddress]{revtex4-2}

\usepackage{graphicx,color}
\usepackage{epsfig}
\usepackage{dcolumn}
\usepackage{amsmath,amssymb,bm}
\usepackage{hyperref,url}
\usepackage[noabbrev]{cleveref}
\usepackage{CJK}
\usepackage{lineno}
\usepackage[english]{babel}
\usepackage[T1]{fontenc}
\usepackage{pifont}
\usepackage{amsfonts}
\usepackage{wasysym}
\usepackage{amsbsy}
\usepackage{psfrag}
\usepackage{color}
\usepackage{multirow}
\usepackage{cases}
\usepackage{stackengine}
\usepackage{float}
\usepackage{blkarray}
\usepackage{cancel}
\usepackage{mathrsfs}
\usepackage{empheq}
\usepackage{mathptmx} 

\DeclareMathAlphabet{\mathbfsf}{\encodingdefault}{\sfdefault}{bx}{sl}

\providecommand\bnabla{\boldsymbol{\nabla}}

\newcommand{\p}[1]{\partial}
\newcommand{\dd}[2] { \frac{\mathrm{d} #1}{\mathrm{d} #2} }

\newcommand{\com}[1]{\textcolor{black}{#1}}

\begin{document}
\preprint{Revised version submitted to Phys. Rev. Lett.}

\title{Universal Free-Fall Law for Liquid Jets under Fully Developed Injection Conditions}

\author{M. Beneitez}
\affiliation{DAMTP, Centre for Mathematical Sciences, Wilberforce Road, Cambridge CB3 0WA, United Kingdom}

\author{D. Moreno-Boza}
\affiliation{Grupo de Mec\'anica de Fluidos, Departamento de Ingenier\'ia T\'ermica y de Fluidos, Universidad Carlos III de Madrid. Avda. de la Universidad 30, 28911, Legan\'es, Madrid, Spain.}

\author{A. Sevilla}
\email{Corresponding author: alejandro.sevilla@uc3m.es}
\affiliation{Grupo de Mec\'anica de Fluidos, Departamento de Ingenier\'ia T\'ermica y de Fluidos, Universidad Carlos III de Madrid. Avda. de la Universidad 30, 28911, Legan\'es, Madrid, Spain.}

\begin{abstract}
We show that vertical slender jets of liquid injected in air with a fully-developed outlet velocity profile have a universal shape in the common case in which the viscous force is much smaller than the gravitational force. The theory of ideal flows with vorticity provides an analytical solution that, under negligible surface tension forces, predicts $R_j(Z)=[(1+Z/4)^{1/2}-(Z/4)^{1/2}]^{1/2}$, where $R_j$ is the jet radius scaled with the injector radius and $Z$ is the vertical distance scaled with the gravitational length, $l_g=u_o^2/2g$, where $u_o$ is the mean velocity at the injector outlet and $g$ is the gravitational acceleration. In contrast with Mariotte's law, $R_j=(1+Z)^{-1/4}$, previously reported experiments employing long injectors collapse almost perfectly onto the new solution.
\end{abstract}

\date{\today}

\maketitle


\emph{Introduction}.-- The structure, dynamics and rupture of freely falling liquid jets in air have been the subject of scientific inquiry for centuries~\citep{Eggers1997,EyV}. Although early insightful observations appear in Da Vinci's Codex Leicester~\citep{daVinci}, the first quantitative theory of falling liquid columns is due to Mariotte~\citep{Mariotte}, who made use of Torricelli's free-fall law $u_j(z)=(u_o^2+2gz)^{1/2}$, where $u_j(z)$ is the axial velocity of the liquid, assumed to be uniform in the radial direction, $u_o$ is the injection velocity, $z$ is the vertical distance measured downwards from the release height, and $g$ is the gravitational acceleration~\citep{Torricelli}. When combined with the liquid continuity equation, $u_j(z) r_j(z)^2=u_o r_o^2$, where $r_o$ and $r_j(z)$ are the injector and jet radii, respectively, the dimensionless jet radius $R_j(z)=r_j(z)/r_o$ obeys Mariotte's law, $R_j=(1+2gz/u_o^2)^{-1/4}$. The convective acceleration of the liquid due to gravity, $\bm{v}\cdot\bnabla \bm{v}\sim u_o^2/l_g \sim g$ defines a gravitational length $l_g=u_o^2/2g$ such that 
\begin{equation}\label{eq:mariotte}
R_j(Z)=(1+Z)^{-1/4},
\end{equation}
in terms of the rescaled height $Z=z/l_g=z/(Fr\,r_o)$, where $Fr=l_g/r_o=u_o^2/(2gr_o)$ is the Froude number.

Mariotte's law~\eqref{eq:mariotte} provides a highly accurate description of the asymptotic structure of free liquid jets for $Z\gg 1$, a limit in which all the forces other than liquid inertia and gravity become negligibly small~\citep{SenchenkoBohr}. These additional forces include, among others, surface tension~\citep{geer83}, viscous stresses~\citep{EggersDupont,GyC}, aerodynamic forces~\citep{Weber1931,Sterling1975}, and the radial diffusion of liquid momentum that takes place when the initial velocity profile is non-uniform~\citep{DudayVrentas,Goren66,GorenyWronski,OguzRelaxation,franceses,sevilla2011}. Here we focus on the latter viscous relaxation effect, which is particularly important in actual engineering flows due to the fact that most liquid injection devices involve the development of viscous boundary layers at the inner injector wall, with the fully developed Poiseuille profile as the canonical example of non-uniform exit conditions. The downstream diffusion of the outlet vorticity in the free-jet region depends on the Reynolds number, $Re=u_o r_o/\nu$, where $\nu$ is the kinematic viscosity of the liquid. When $Re\lesssim 1$ the velocity profile becomes uniform within a distance $l_{\nu}\sim r_o$ from the outlet. In contrast, when $Re\gg 1$, the viscous development length $l_{\nu}\sim u_o r_o^2/\nu= Re\,r_o\gg r_o$, as deduced by balancing the convective acceleration, $\bm{v}\cdot\bnabla \bm{v}\sim u_o^2/l_{\nu}$, with the radial diffusion of momentum, $\nu\bnabla^2\bm{v} \sim \nu u_o/r_o^2$. The structure of the relaxing free jet is thus controlled by the ratio $l_{\nu}/l_g \sim 2 g r_o^2/(\nu u_o)=G=Re/Fr$. The gravity number $G$ allows to distinguish two relevant limiting cases: when $G\ll 1$, the viscous relaxation takes place under almost buoyancy-free conditions~\citep{DudayVrentas,GorenyWronski}, and the gravitational thinning occurs far downstream, where the velocity profile is almost uniform. In the opposite limit $G\gg 1$, the viscous diffusion of momentum is negligible along the initial gravitational thinning region.

Here we provide a first-principles description of the jet structure in the frequently realized limit $G\gg 1$ assuming, for simplicity, the slenderness \com{hypothesis $Fr\gg 1$}. To that end, advantage is taken of the fact that viscous forces can be neglected, in a first approximation, within the gravitational stretching region $0<z\lesssim l_g\ll l_{\nu}$. Under these conditions, the theory of ideal flows with vorticity~\citep{batchelor2000} is the appropriate framework to describe the structure of the falling jet. 


\begin{figure*}
    \centering
    \includegraphics[width=0.7\textwidth]{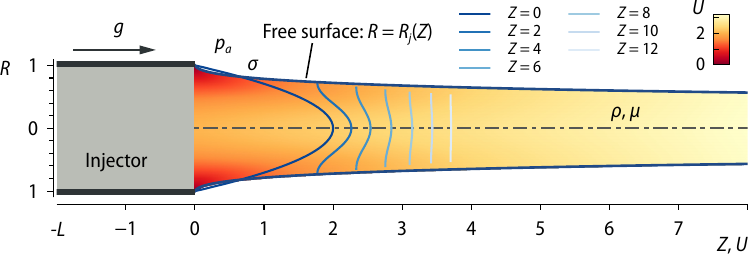}
    \caption{Sample numerical jet with parameter definitions and cylindrical system of coordinates for $Re = 400, Fr = We = 10$. Accompanying axial velocity profiles are represented for selected axial locations.}
    \label{fig:sketch}
\end{figure*}

\emph{Theory of inviscid vortical jets}.-- Consider a liquid jet emerging from a circular tube of radius $r_o$, and injected downwards into a passive gaseous atmosphere at constant pressure $p_a$ under the presence of a gravitational acceleration $g$. An example of the resulting liquid-air interface, of constant surface tension coefficient $\sigma$, can be observed in Fig.~\ref{fig:sketch}, where the main variables used in the analysis are also indicated. We describe the jet with use made of a cylindrical coordinate system $(z,r,\theta)$ with origin at the center of the injector exit and the axial coordinate $z$ pointing in the direction of gravity. The liquid velocity field, $\bm{v}=u(r,z)\bm{e}_z+v(r,z)\bm{e}_r$, is assumed incompressible, inviscid and axisymmetric, and thus satisfies the momentum equation
\begin{equation}\label{eq:cm}
    \bnabla \left(\dfrac{u^2+v^2}{2}+\dfrac{p}{\rho} -gz \right) = \bm{v} \wedge \left(\bnabla \wedge \bm{v}\right),
\end{equation}
where $p$ is the liquid pressure and $|\bm{v}|^2/2=(u^2+v^2)/2$ is the kinetic energy per unit mass. The projection of Eq.~\eqref{eq:cm} along the jet streamlines yields Bernoulli's equation
\begin{equation}\label{eq:cmStream}
   \dfrac{u^2+v^2}{2}+\dfrac{p}{\rho} -gz = C_l,
\end{equation}
indicating that the total pressure remains constant along each streamline, but is different for different streamlines. Indeed, in contrast with the case of an irrotational flow, here the value of $C_l$ depends on the streamline due to the fact that $\bnabla \wedge \bm{v} \neq 0$. The continuity equation $\partial_z( r u)+\partial_r(r v)=0$ provides the estimate $v_c\sim u_o r_o/l_g=u_o/Fr \ll u_o$ for the characteristic radial velocity, whereby the kinetic energy $|\bm{v}|^2/2 \simeq u^2/2$ with small $O(Fr^{-2})$ relative errors. In addition, it is useful to introduce a stream function $\psi$ defined by $ru=\partial_r \psi$ and $rv=-\partial_z \psi$, satisfying the continuity equation, such that Eq.~\eqref{eq:cmStream} reads
\begin{equation} \label{eq:ideal}
    U^2(\Psi,Z) = U^2(\Psi,0) + Z + P(\Psi,0) - P(\Psi,Z),
\end{equation}
in its simplified dimensionless form, where $U=u/u_o=R^{-1}{\rm d}\Psi/{\rm d}R$, $\Psi=\psi/(u_o r_o^2)$, $P=2p/(\rho u_o^2)$,  $R=r/r_o$ and $Z=z/l_g$. To close the model, an appropriate expression for the pressure $P$ is needed. To that end, use is made of the normal stress balance at the free surface, $p-p_a=\sigma \mathcal{C}$, where $\mathcal{C}$ is twice the mean curvature of the free surface. The slenderness assumption $Fr\gg 1$ allows to simplify the curvature, $\mathcal{C}=r_j^{-1}$, and therefore the normal stress balance may be reduced to $p=p_a+\sigma/r_j$, with relative errors $O(Fr^{-2})\ll 1$. In addition, the radial pressure variations inside the liquid, $\Delta_r p$, can also be neglected with the same relative error. Indeed, the radial momentum equation provides the estimate $\Delta_r p\sim \rho u_o v_c r_o/l_g \sim \rho u_o^2\,Fr^{-2}\ll \rho u_o^2$. With these simplifications, the non-dimensional pressure difference between the jet exit, $Z=0$, and a generic downstream station $Z$, reads $P(0)-P(Z) = We^{-1}(1-1/R_j)$, where $We = \rho u_o^2 r_o/(2\sigma)$ is the Weber number, which is the only parameter governing the structure of the vortical ideal jet.

Assuming that the outlet velocity profile is fully developed\com{~\footnote{\com{The theory can be readily applied to arbitrary outlet velocity profiles but must, in general, be integrated numerically.}}}, $U(R,0)=2(1-R^2)$, the stream function at $Z=0$ is obtained by radial integration as $\Psi = R^2(1-R^2/2)$, which can be inverted to yield $R^2=1-(1-2\Psi)^{1/2}$. It follows then that $U^2(\Psi,0) = 4(1-2\Psi)$, whereby Eq.~\eqref{eq:ideal} reduces to the ordinary differential equation\com{~\footnote{\com{In the non-slender case with $Fr\lesssim 1$ the radial and axial velocities are comparable, and the capillary pressure must include the contribution of the axial curvature. The mathematical model can then be formulated as an elliptic free-boundary problem whose analysis is much more involved than the slender approximation employed herein.}}
\begin{equation}\label{eq:edo}
    \dfrac{1}{R^2}\left(\dd{\Psi}{R}\right)^2 = 4(1-2\Psi) + F(Z),
\end{equation}
for the function $\Psi(R;Z,R_j,We)$, where $F(Z)=Z-\frac{1}{We}\,\frac{1-R_j}{R_j}$, and $Z,\,R_j(Z)$ play the role of parameters in the slender approximation employed here. The integration of Eq.~\eqref{eq:edo} by separation of variables yields
\begin{equation}\label{eq:edo_int}
    \int_0^{1/2}\frac{{\rm d}\Psi}{\sqrt{4(1-2\Psi) + F(Z)}} = \frac{R_j^2}{2}
\end{equation}
where application has been made of the symmetry boundary condition at the axis, $\Psi(R=0)=0$, and of the conservation of the flow rate, $\Psi(R=R_j)=1/2$. A straightforward integration of Eq.~\eqref{eq:edo_int} finally provides the octic}
\begin{equation} \label{eq:sol}
    R_j^8-(2+Z+We^{-1})R_j^4+We^{-1}R_j^3+1=0,
\end{equation}
which, for $We\gg 1$, has the explicit solution
\begin{equation}\label{eq:Rj(Z)}
R_j(Z) = \sqrt{\sqrt{1+Z/4}-\sqrt{Z/4}},
\end{equation}
to be compared with Mariotte's law~\eqref{eq:mariotte}. The far-field shape of the jet is found by expanding Eq.~\eqref{eq:Rj(Z)}, to yield $R_j=Z^{-1/4}-Z^{-5/4}/2+O(Z^{-9/4})$ for $Z\gg 1$, while Eq.~\eqref{eq:mariotte} has the large-$Z$ expansion $Z^{-1/4}-Z^{-5/4}/4+O(Z^{-9/4})$. Thus, both solutions converge to the same asymptotic free-fall law $R_j \to Z^{-1/4}$ as $Z \to \infty$. In contrast, the shapes predicted near the injector are completely different. Indeed, Eq.~\eqref{eq:Rj(Z)} provides $R_j=1-Z^{1/2}/4+Z/32+O(Z^{3/2})$ for $Z\ll 1$, while Mariotte's law~\eqref{eq:mariotte} yields $1-Z/4+5Z^{2}/32+O(Z^3)$. It is thus deduced that the initial contraction of the jet is much more abrupt for the jet with initial Poiseuille profile than for the uniform jet, as a consequence of the smaller velocity of the fluid particles near the free surface.

\begin{figure*}[t]
  \includegraphics[width=0.75\textwidth]{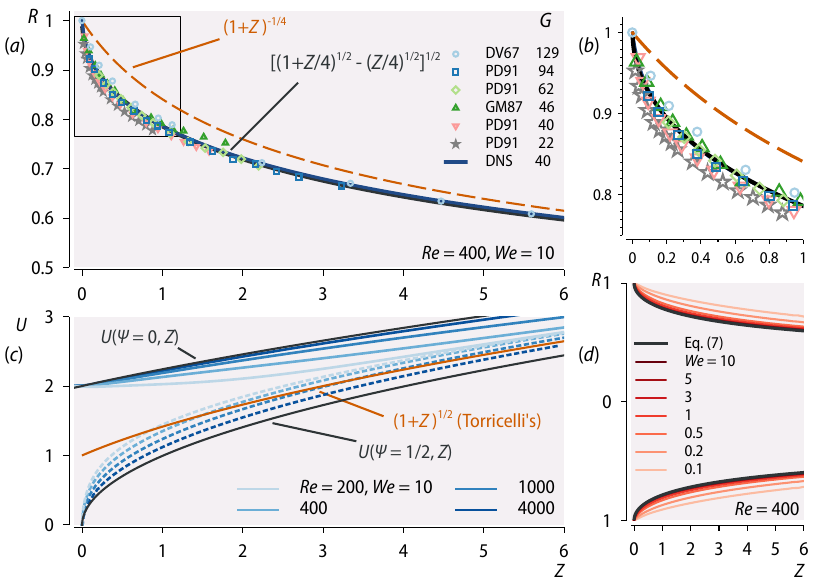}
  \caption{(a) and (b) Comparison of the jet shapes given by Mariotte's law~\eqref{eq:mariotte}, and the new solution~\eqref{eq:Rj(Z)}, with experimental data for $G\gg 1$ and $We\gtrsim 1$ extracted from \citet{DudayVrentas} (DV67),~\citet{franceses} (PD91) and~\citet{gonzalez1987} (GM87). The free surface from a steady numerical simulation of a jet for $Re = 400$, $Fr = We = 10$ is also presented for comparison. (c) Downstream evolution of the jet velocity at the axis (solid lines) and at the free surface (dashed lines) obtained from numerical simulations at different values of $Re$, together with the solution of the ideal vortical flow~\eqref{eq:ideal} and Mariotte's free-fall law~\eqref{eq:mariotte}. (d) Influence of capillarity on the shape of the jet for several $We$ for $Re = 400$, $Fr = 10$, showing that surface-tension effects are negligible for $We\gtrsim 3$.\label{fig:univ}}
\end{figure*}

\emph{Effect of surface tension}.-- Weak surface-tension effects can be rationalized by means of a perturbation expansion of Eq.~\eqref{eq:sol} in powers of $We^{-1}$, to yield $R_j(Z;We)=[(1+Z/4)^{1/2}-(Z/4)^{1/2}]^{1/2}\left[1+We^{-1}f(Z)\right]$. 
The function $f(Z)$ has a maximum value of 0.032, whereby the relative errors introduced by Eq.~\eqref{eq:Rj(Z)} compared with Eq.~\eqref{eq:sol} are smaller than 1\% for $We\gtrsim 3$. Similarly, surface tension effects can also be introduced in Mariotte's solution~\eqref{eq:mariotte} by considering the capillary pressure jump in the slender approximation, to yield $U_j^2=1+Z-We^{-1}(1-R_j)/R_j$, leading to the quartic 
$(1+Z+We^{-1})R_j^4-We^{-1}R_j^3-1=0$
which has a large-$We$ expansion $R_j=(1+Z)^{-1/4}[1+We^{-1}s(Z)]$, where the function 
$s(Z)$ has a single maximum $s(Z=175/81)=27/1024\approx 0.0264$, so that the relative error introduced by Eq.~\eqref{eq:mariotte} is smaller than 1\% for $We\gtrsim 2$. These conclusions are confirmed by the results of Fig.~\ref{fig:univ}(d), where full Navier-Stokes simulations are compared with Eq.~\eqref{eq:Rj(Z)} for several values of $We$.

\emph{Numerical simulations}.-- The steady axisymmetric Navier-Stokes equations were integrated numerically using Comsol. Finite elements are employed to discretize the deformable domain, whose displacement field is computed by solving a Laplace equation within an arbitrary Lagrangian-Eulerian (ALE) framework. Conversely to what has been presented above, for this section we will take $r_o$ as the characteristic length scale for both axial and radial directions so that the dimensionless equations of motion read 
\begin{equation} \label{eq:NS}
    \bnabla \cdot \bm{v} = 0, \quad \bm{v} \cdot \bnabla \bm{v} = - \bnabla P/2 + \bnabla \cdot \bar{\bar{\tau}} +  (2 Fr)^{-1} \, \bm{e}_z,
\end{equation} where $\bar{\bar{\tau}} = (\bnabla \bm{v} + \bnabla\bm{v}^T)/Re$ is the viscous stress tensor, and the necessary boundary conditions are $U(-L/r_o,R) - 2(1-R^2) = V(-L/r_o,R) = 0$ (Poiseuille flow prescribed at the upstream boundary of an injector of length $L$), $U(Z,1)= V(Z,1) = 0$ (no-slip at the injector inner wall) for $Z \leq 0$, $V(Z,0) = 0$ (radial symmetry), and $-(P/2)\bm{n} + \bar{\bar{\tau}} \cdot \bm{n} = - (2 We)^{-1} \bm{n} \bnabla \cdot \bm{n}$ (stress balance) at the free surface $R = R_j(Z)$, which needs to be determined as part of the solution, where $\bm{n}$ is the outwards normal to the interface. To that end, the vertical displacement of the mesh at the free-surface boundary is left as a Lagrange multiplier that enforces the kinematic condition $\bm{v} \cdot \bm{n} = 0$ or, equivalently, $U \, \partial{R_j}/\partial{Z} = V$. A stress-free boundary condition at $Z \gg 1$ completes the description of the numerical set-up. The solution is found employing a root-finding algorithm with a prescribed normalized tolerance not greater than $10^{-4}$. The prescription of Eq.~\eqref{eq:Rj(Z)} as initial seed for the free-surface shape was seen to facilitate the convergence greatly. Several other initial guesses were also employed to assess numerical independence. An example of the numerical jet so obtained is presented in Fig.~\ref{fig:sketch} for $Re = 400$, $Fr = We = 10$.

\emph{Comparison with experiments}.-- Due to the fundamental and applied interest of free liquid jets, a large number of experimental studies can be found in the literature covering many different aspects of the jet structure, instability and break-up~\citep{Eggers1997,EyV}. Among these studies, several ones were devoted to the generation of slender gravity-dominated jets emerging from long injectors fulfilling the hypotheses behind our analysis. Figure~\ref{fig:univ}(a,b) shows the free-surface shape extracted from several available experiments (symbols), a Navier-Stokes simulation at $Re=400$ and $Fr=We=10$, together with the classical free-fall solution given by Eq.~\eqref{eq:mariotte} and our parameter-free, leading-order solution~\eqref{eq:Rj(Z)}. In all cases the experimental jet shapes are much closer to our free-fall law~\eqref{eq:Rj(Z)} than to Mariotte's equation~\eqref{eq:mariotte}, most notably in the region close to the injector outlet, as illustrated in Fig.~\ref{fig:univ}(b). The slight deviations from Eq.~\eqref{eq:Rj(Z)} observed in Fig.~\ref{fig:univ} can be attributed to the finite values of $G$ in the experiments, leading to a non-negligible radial diffusion of momentum across the jet core (see e.g. the star symbols extracted from Ref.~\cite{franceses}). The downstream evolution of the axial velocity at the jet centerline and at the free surface provided by the numerical simulation is represented in Fig.~\ref{fig:univ}(c) for several values of $Re$, together with the predictions of~\eqref{eq:ideal} for the ideal vortical flow, as well as Torricelli's free-fall law. These results clearly show that the Navier-Stokes flow converges to the ideal flow solution as the Reynolds number is increased.

\emph{Concluding remarks}.-- Our new analytical solution~\eqref{eq:Rj(Z)} for the steady structure of slender laminar liquid jets with fully-developed outlet velocity profiles provides much better agreement with experiments than the routinely used Mariotte's law~\eqref{eq:mariotte}. This comparison unambiguously demonstrates that, while Mariotte's solution fails to describe liquid jets with vorticity at large Reynolds numbers, the new theory predicts a universal shape function that collapses all the experiments onto a single master curve.

The hypotheses behind the theory, namely $Fr\gg 1$ and $Re\gg Fr$, clearly need some justification. To that end, it proves convenient to introduce the Morton number \mbox{$Mo=g\mu^4/\rho\sigma^3$}, and the Bond number \mbox{$Bo=\rho g r_o^2/\sigma$} as auxiliary dimensionless parameters, in terms of which \mbox{$Fr=We/Bo$} and \mbox{$G=Re/Fr=\sqrt{2}Mo^{-1/4}Bo^{5/4}We^{-1/2}$}. Large values of the Froude number require that $Bo\ll We$, a condition that is accomplished for injectors of sufficiently small diameter in the jetting regime, which requires $We\gtrsim 1$ for $Bo\ll 1$~\citep{Clanet1999,rubio2013}. In contrast, the condition $G\gg 1$ depends crucially on the viscosity of the working liquid. For instance, the value of $Mo\approx 2.5\times 10^{-11}$ in the case of water at room temperature, in which case $G\approx 631 Bo^{5/4}We^{-1/2}$, and the validity condition writes $We\ll 4\times 10^5 Bo^{5/2}$, which is accomplished under the Rayleigh break-up regime for any realistic injector diameter~\citep{Gordillo2001}. Although similar conclusions hold for any low-viscosity liquid, the validity of our theory breaks down for liquids of sufficiently large viscosity, for which either full Navier-Stokes simulations or simplified one-dimensional descriptions may be used instead~\citep{Eggers1997,EyV}.


\begin{acknowledgments}
Useful discussions with A. Mart\'inez-Calvo and M. Rubio-Rubio are gratefully acknowledged. The authors thank the Spanish MCIU-Agencia Estatal de Investigaci\'on through project PID2020-115655GB-C22, partly financed through FEDER European funds.
\end{acknowledgments}

\end{document}